\documentstyle[twocolumn,floats,ifthen,aps,prl]{revtex}

\begin{document}
\draft
\title{Possible cooling by resonant Fowler-Nordheim emission}
\author{Alexander N. Korotkov and Konstantin K. Likharev}
\address{Department of Physics, State University of New York at Stony Brook, 
Stony Brook, NY 11794-3800 }
\date{\today}

\maketitle

\begin{abstract}
A new method of electronic refrigeration based on resonant Fowler-Nordheim
emission is proposed and analyzed. In this method, a bulk emitter is covered
with a-few-nm-thick film of a widegap semiconductor, creating an
intermediate step between electron energies in the emitter and in vacuum. An
external electric field tilts this potential profile, forming a quantum
well, and hence 2D electron subbands at the semiconductor-vacuum boundary.
Alignment of the lowest subband with the energy levels of the hottest
electrons of the emitter (a few $k_{B}T$ above its Fermi level) leads to a
resonant, selective emission of these electrons, providing emitter cooling.
Calculations show that cooling power as high as 10$^{4}$ W/cm$^{2}$ (at 300
K), and temperatures down to 10 K may be achieved using this effect.
\end{abstract}



\narrowtext 

\vspace{1cm}

The idea of using thermionic transport of electrons over an energy barrier
for cooling has been repeatedly discussed in the literature (see, e.g., \cite
{Mahan,Shakouri,Mahan-2}). If the barrier height is a few times the thermal
spread $k_{B}T$, the thermionic current may be quite substantial, with only
the hot fraction of electrons being removed from the conductor.
Unfortunately, the practical implementation of this idea runs into problems.

A barrier of the necessary height ($\sim $200 meV for 300 K, and
proportionally lower for lower temperatures) may be readily implemented in
solid state structures, in particular using composite semiconductors with
the necessary conduction band edge offset. However, even if the barriers are
relatively thick, the back flow of heat to the cooled conductor is
prohibitively high \cite{Mahan,Shakouri,Mahan-2}; multilayer structures
proposed to overcome this effect \cite{Mahan-2} seem very complex and
promise only a little cooling power. (Only at very low, millikelvin, 
temperatures
where electron-phonon coupling is very weak, has efficient cooling been
demonstrated using thermionic transfer over the superconductor energy gap 
\cite{Nahum,Levio}.)

Even a very narrow (submicron) vacuum gap can effectively quench the back
heat flow, reducing it to radiation-limited levels of the order of 0.1 
W/cm$^{2}$ (at 300 K). Unfortunately, in this case the energy barrier height 
is determined by the conductor workfunction which is too high ($\gg k_{B}T$)
for most materials. A natural way to enforce electron transfer through a
relatively high barrier is to apply a strong electric field ($\sim$ 10
MV/cm), inducing Fowler-Nordheim tunneling through the barrier. However, in
typical situations the tunneling through the initially uniform barrier pulls
out electrons within a relatively broad energy range that results in heating
rather than in cooling (the ``Nottingham effect'' \cite{Nottingham}).

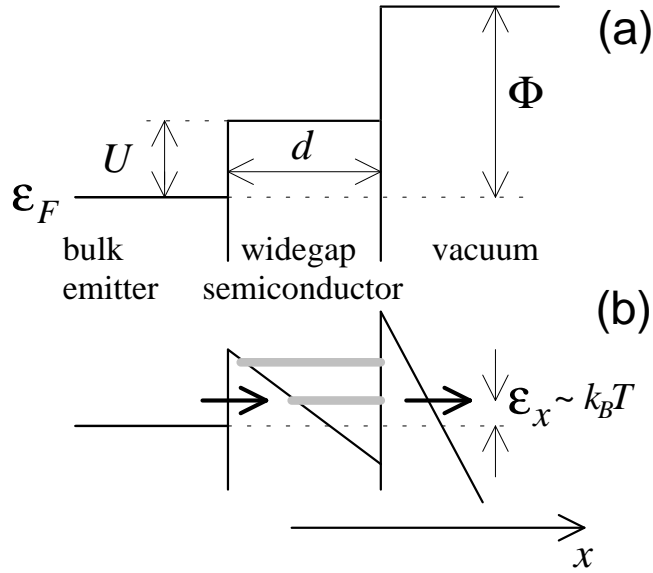
\begin{figure}[tbp]
\caption{The energy diagram of the proposed device: (a) in absence of bias
and (b) at finite electric field. Resonant tunneling via quantized levels
above the Fermi energy removes the hot fraction of electrons, thus cooling
the emitter. }
\label{fig1}
\end{figure}

We propose to limit the energy range of transferred electrons using resonant
tunneling in a simple structure (Fig.\ \ref{fig1}) where the bulk emitter (a
metal or a heavily doped semiconductor) is covered with a thin (a-few-nm)
layer of a widegap semiconductor. While at zero voltage the electron
potential energy profile of this structure has two steps (Fig.\ 
\ref{fig1}(a)), its tilting by the applied electric field creates 
a triangular-shape
potential well at the semiconductor film surface (Fig.\ \ref{fig1}(b)).
Quantization of the energy of electron motion perpendicular to the surface
results in discrete levels (subbands for the full energy) localized at the
surface. If the electric field aligns the lowest subband with energy levels
of the hottest electrons in the emitter (a few $k_{B}T$ above the Fermi
level), resonant tunneling of these electrons to vacuum may lead to very
efficient heat removal, and hence to emitter cooling \cite{crest}.

Our proposal hinges on several ideas put forward earlier. First of all,
numerous experiments indicate that Fowler-Nordheim emission from bulk
cathodes is frequently enhanced by resonant tunneling through localized
surface states arising from its unintentional contamination (see, e.g.,
Refs.\ \cite{Gadzuk,Binh}). Cooling of the nanoclusters using this effect
was proposed in Ref.\ \cite{Purcell}. (Cooling of 2D electron gas based on
the resonant tunneling through specially fabricated quantum dots was
proposed even earlier\ \cite{Edwards}). However, to extend cooling to
macroscopic objects, a large number of surface nanoparticles should be used
in a single device. In this case, unavoidable spread of the size and shape
of these particles would result in fluctuations of the resonant level
positions, preventing their proper alignment with the Fermi level of the
emitter, unless nanoscale fabrication with atomic precision is used - a very
distant prospect indeed. In contrast, our suggestion involves only planar
structures and does not require nanofabrication.

Concerning planar structures, Fowler-Nordheim tunneling via the resonant
subbands was predicted long ago \cite{Gundlach} and then observed in several
systems \cite{Maserjian,Pettersson,Lewicki,Maserjian-2,Hickmont,Mimura}. The
emission via resonant subbands at the outer surface of a widegap 
semiconductor in a strong electric field was predicted in Ref.\ \cite
{Litovchenko}. The Fowler-Nordheim emission coupled to the electron
resonance in the vacuum gap was considered in Ref.\ \cite{Vatannia} (the
implementation of this effect would, however, require an impractically fine
uniformity of the gap). However, the possibility of heat removal was not
mentioned in any of these publications.

The objective of this work was a quantitative analysis of the cooling effect
in the system shown in Fig.\ \ref{fig1}, using a simple but natural model.
First of all, we assume the interfaces to be perfectly plane. In this case
the electron motion in the direction of tunneling ($x$-axis) and in the
perpendicular direction (along the interfaces) are separated. Neglecting
band bending due to quantum well charging (i.e. assuming its shape to be
triangular) we have the following well known result for the electron
eigenenergies (see, e.g., \cite{wells}): 
\begin{eqnarray}
&&{\cal E}={\cal E}_{x}+{\cal E}_{\perp },\,\,\,{\cal E}_{x}=U-eEd+
{\cal E}_{n}, \\
&&{\cal E}_{n}=(-a_{n})\left( e^{2}E^{2}\hbar ^{2}/2m\right) ^{1/3}
\end{eqnarray}
where all energies are relative to the emitter Fermi level, $U$ is the
initial energy step (Fig.\ \ref{fig1}), $E$ is the electric field in the
film, $d$ is the film thickness, ${\cal E}_{\perp }=\hbar ^{2}k_{\perp
}^{2}/2m$, $m$ is the electron effective mass in the conduction band of the
film, and $a_{n}$ is the sequence of Airy function zeros: $a_{0}=-2.34$, 
$a_{1}=-4.09,\ldots ,a_{n}\rightarrow -[3\pi (n+3/4)/2]^{2/3}$.

In absence of energy relaxation, the level filling probability $p=p_{n}
({\cal E}_{\perp })$ may be found from the stationary solution of the usual
master equation, giving $p=f \gamma _{L}/(\gamma _{L}+\gamma _{R}),$ where 
$f=f({\cal E)}$ is the Fermi distribution of the emitter electrons, and 
$\gamma _{L}$ and $\gamma _{R}$ are the rates of electron escape from the
quantum well into conductor and into vacuum, respectively. These rates may
be calculated as $\gamma _{L,R}=\nu D_{L,R}$, where $\nu $ is the ``attempt
frequency'', $\nu =[2\int dx/v(x)]^{-1}={\cal E}_n/2\hbar |a_n|^{3/2}$, and 
$D_{L,R}$ are transparencies of the left and right triangular barriers [Fig.\ 
\ref{fig1}(b)]. Within the WKB approximation (neglecting the image charge
effects), 
\begin{eqnarray}
\ln D_{L} &=&-\frac{4\sqrt{2m}}{3eE\hbar }\,(eEd-{\cal E}_{n})^{3/2}, \\
\ln D_{R} &=&-\frac{4\sqrt{2m_{0}}}{3eE_{0}\hbar }\,(\Phi -U- 
{\cal E}_{n}-\Delta {\cal E})^{3/2}.  \label{D_R}
\end{eqnarray}
Here the shift $\Delta {\cal E}=(\hbar ^{2}k_{\perp
}^{2}/2)(m^{-1}-m_{0}^{-1})$ is due to the possible difference between $m$
and the electron mass $m_0$ in vacuum, $\Phi $ is the work function of the
bulk emitter, and $E_{0}$ is the electric field in vacuum. The relation
between this field and $E$ includes the 2D charge density $\sigma $ of the
electrons accumulated in the well: 
\begin{equation}
\epsilon _{0}E_{0}=\epsilon \epsilon _{0}E+\sigma ,  \label{field}
\end{equation}
($\epsilon $ is the dielectric constant of the semiconductor film). The
charge density $\sigma $, as well as the resonant current density $j$ and
thermal flow $q,$ may be calculated as 
\begin{equation}
\sigma =\sum_{n,{\cal E}_{\perp }}e\,p\,,\text{ \ }j=\sum_{n,{\cal E}_{\perp
}}e\,\gamma _{R}p,\text{ \ }q=\sum_{n,{\cal E}_{\perp }}{\cal E} \gamma
_{R}p.  \label{sigma}
\end{equation}

When the quantized level is above the emitter Fermi level, the typical
spread of ${\cal E}_{\perp }$ for the electrons in the subband is of the
order of temperature $T$ (here and below measured in energy units). Hence,
assuming that the barriers are much higher than $T$, we can neglect the term 
$\Delta {\cal E}$ in Eq.\ (\ref{D_R}). Then integrating over 
${\cal E}_{\perp }$, we get: 
\begin{eqnarray}
&&j=e\rho \sum_{n}\,T\ln (1+e^{-{\cal E}_{x}/T})\,\gamma _{L}\gamma
_{R}/(\gamma _{L}+\gamma _{R}),  \label{j} \\
q &=&\rho \sum_{n}\Big[{\cal E}_{x}T\ln (1+e^{-{\cal E}_{x}/T})+\frac{T^{2}}
{2}[\ln (1+e^{-{\cal E}_{x}/T})]^{2}  \nonumber \\
&&+T^{2}\mbox{Li}_{2}[(1+e^{{\cal E}_{x}/T})^{-1}]\Big]\,\gamma _{L}\gamma
_{R}/(\gamma _{L}+\gamma _{R}),  \label{q}
\end{eqnarray}
where $\mbox{Li}_{2}(z)=\sum_{k=1}^{\infty }z^{k}/k^{2}$ is the dilogarithm
function and $\rho =m/\pi \hbar ^{2}$ is the 2D density of states per unit
area.

Equations (\ref{j}), (\ref{q}) do not include the components of current 
($j^{\prime }$) and heat flow ($q^{\prime }$) 
which are due to nonresonant, direct tunneling through the
complete energy barrier. For this process, the barrier transparency may be
calculated as $D=D_{L}D_{R}$. A standard WKB calculation yields: 
\begin{eqnarray}
&&j^{\prime }=\frac{e{\cal E}_{0}^{2}m_{c}}{2\pi ^{2}\hbar ^{3}K} 
\,D_{0}^{L}D_{0}^{R}\,\frac{t}{\sin t}\,,\,\,\,\,K=\frac{m_{c}{\cal E}_{0}}
{m{\cal E}_{0}^{L}}+\frac{m_{c}{\cal E}_{0}}{m_{0}{\cal E}_{0}^{R}}  
	\label{j'} \\
&&q^{\prime }=-\frac{{\cal E}_{0}^{3}m_{c}}{2\pi ^{2}\hbar ^{3}K}
\,D_{0}^{L}D_{0}^{R}\,\frac{t^{2}\cos t}{(\sin t)^{2}}\,,\,\,\,\,\,\,\,t=\pi
T/{\cal E}_{0},
	\label{q'}	\end{eqnarray}
where $1/{\cal E}_{0}\equiv d(-\ln D)/d{\cal E=}1/{\cal E}_{0}^{L}+1/
{\cal E}_{0}^{R}$, 
\begin{eqnarray*} 
{\cal E}_{0}^{L} &=&e\hbar E/2(2m)^{1/2}[U^{1/2}-\max (0,U-eEd)^{1/2}], \\
{\cal E}_{0}^{R} &=&e\hbar E_{0}/2(2m_{0})^{1/2}(\Phi -eEd)^{1/2}, \\
\ln D_{0}^{L} &=&-\frac{4(2m)^{1/2}}{3eE\hbar }\,
[(U-{\cal E}_{x})^{3/2}-\max (0,U-{\cal E}_{x}-eEd)^{3/2}], \\
\ln D_{0}^{R} &=&-\frac{4(2m_{0})^{1/2}}{3eE_{0}\hbar }\,\max (0,\Phi -eEd-
{\cal E}_{x})^{3/2}.
\end{eqnarray*}
(Notice that these formulas may be used even if ${\cal E}_{x}>U-eEd$. The
exact calculation in this case would give an extra factor on the order of
unity, however, it can be neglected within the accuracy of WKB
approximation.) The well-known factor \cite{Lenzlinger} $t/\sin t$ shows
that our approximation, based on the linear expansion of $\ln D$ near the
Fermi surface, can be used only at $T<{\cal E}_{0}$. (At higher temperatures
the transport at large energies prevails.) At ``low'' temperatures we are
discussing, the nonresonant tunneling always provides heating of the
emitter, although it changes to cooling at the inversion temperature 
$T_{inv}={\cal E}_{0}/2$.

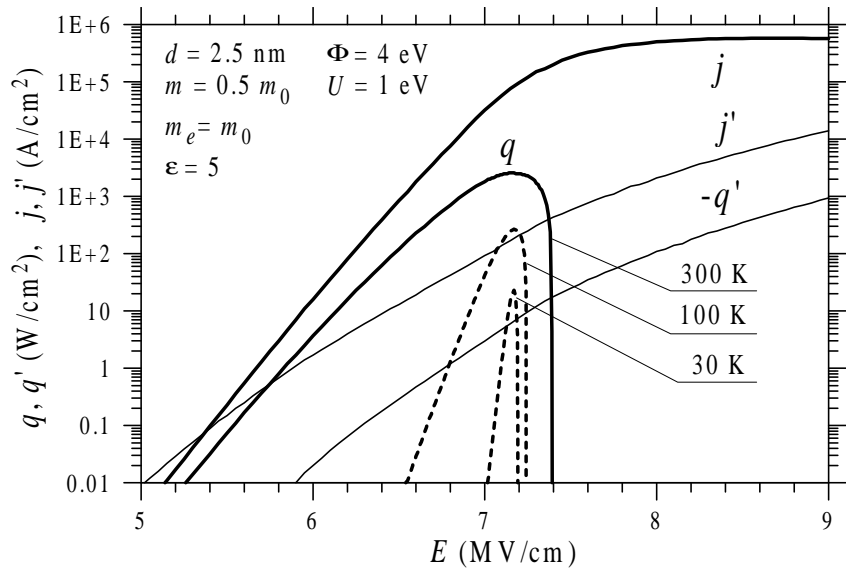
\begin{figure}[tbp]
\caption{Solid lines: the resonant current density $j$, the
corresponding cooling power density $q$, nonresonant current $j^{\prime }$,
and the corresponding heating power $-q^\prime $ as functions of the applied
electric field $E$ for $\Phi =$4 eV, $U=$1 eV, $m=0.5m_{0}$, $m_{c}=m_{0}$, 
$\protect\epsilon =5$, and $d=$2.5 nm, at $T=300$ K. The dashed lines show 
the cooling power $q$ at $T=100$ K and $T=30$ K (the curve for $-q^\prime$
does not change significantly when the temperature is lowered).}
\label{fig2}
\end{figure}

Figure \ref{fig2} shows one of the results of our calculations using Eqs.\  
(\ref{field})-(\ref{q'}). The cooling power first increases exponentially with
the field, because the lowest subband is aligned with more and more
populated hot electron levels, and then drops sharply as soon as the subband
approaches the Fermi level (at larger fields $q$ becomes negative). 
Just before
this drop, the cooling power reaches a maximum, in this case, as high as 
$3\times 10^{3}$ W/cm$^{2}$ at $T=300$ K.

The maximum values of $q,$ as well as the corresponding values of $j$ and 
$q^{\prime }$, for several other parameter sets are listed in Table I. From
Fig.\ \ref{fig2} one can see that the suitable range for the electric field 
$E$ shrinks rapidly as emitter temperature goes down. Nevertheless, our model
indicates that $q+q^{\prime}$ may be positive (i.e., cooling is still
possible) for temperatures as low as 10 K.

Let us discuss how realistic our model is. First of all, Eqs.\ (\ref{sigma})
are valid only if the energy relaxation in the well is much slower than 
$\gamma _{L},$ $\gamma _{R}$. We have also neglected the resonant subband
broadening due to tunneling, but it was monitored through our calculations,
so that the results are presented for only such parameters that the
broadening is negligible in comparison with the important energy scale, $T$.
One more possible source of deviations from our model is electron scattering
in the well and during tunneling. However, these processes cannot lower the
lowest subband deeper into the well, and so can hardly affect the process of
hot electron extraction. Next, we have implicitly assumed that the Fermi
energy of the bulk emitter is much larger than all the considered energies.
If this condition is not satisfied, the results would change, but not
significantly.

Despite the used assumptions, we expect that for smooth films the overall 
accuracy of our 
results for $j$, $q$, $j^{\prime }$, and $q^{\prime }$ for a given applied
field is limited mainly by that of the WKB approximation \cite{Yang} and can
be characterized by a numerical factor of the order of two or three. On the
logarithmic scale at which we are working (Fig.\ \ref{fig2}) this is good
accuracy indeed. Since the results show that the resonant emission cooling
at temperatures above $\sim $100 K may prevail over the nonresonant heating
in a relatively broad range of electric field, and their ratio may be very
high, we are confident that the net cooling of the emitter may be achieved.
However, the estimate of the lowest achievable temperature (10 K)
may be more vulnerable. 

The largest problem we see with the experimental implementation of resonant
emission cooling is the necessary film uniformity. In fact, Table I shows
that at 300 K the effect is stable with respect to substantial ($\sim
20\%$) variations of $d.$ However, to achieve cooling to 100 K, film
thickness variations should not exceed $\sim 4\%$, i.e., about one
monolayer. (Thickness fluctuations require the electric field to be
decreased below the optimal value in order to be sure that we have not
stepped into the heating region on any considerable fraction of the emitter
area.)

While the proposed device potentially offers very large cooling power, its
efficiency (more exactly, the ``coefficient of performance'', COP) may be
relatively low. For our system, COP may be presented as $(q+q^{\prime
})/V(j+j^{\prime })$, where $V$ is the applied voltage. Even if the vacuum
gap $d_{0}$ is as small as 10 nm, the necessary voltage $V=Ed+E_{0}d_{0}$
exceeds 10 V, giving COP about 10$^{-3}$ at $T$ = 300 K. To increase
the COP, the electric field may be provided by a micromachined ``grid''
electrode very close to the surface, followed by another, much more distant
grid at approximately the same voltage, and a collector at lower voltage, so
that electrons are decelerated before the landing (see, e.g., \cite{Ishii}).

To summarize, we have proposed a new method of electronic refrigeration
using the resonant Fowler-Nordheim tunneling in a fairly simple structure.
If the experiment confirms our theory, this device may be invaluable for
heat removal from electronic chips, as well as for the integration of
advanced low-temperature devices with room-temperature circuits.

Useful discussions with D.\ V.\ Averin, H.\ Busta, and R.\ Tsu are gratefully
acknowledged.

\begin{table}[tbp]
\caption{ Maximum cooling flow density $q$ and heating density $-q^{\prime}$,
as well as the corresponding electric field $E$ and resonant electric
current density $j$, for several paratemer sets.}
\label{table1}
\begin{tabular}{|c|c|c|c|c|c|c|c|c|c|}
$\Phi $ & $U$ & $m$ & $\epsilon $ & $d$ & $T$ & $E$ & $j$ & $q$ & 
$-q^{\prime }$ \\ 
eV & eV & $m_{0}$ &  & nm & K & MV/cm & (kA/cm$^{2})$ & W/cm$^{2}$ & 
W/cm$^{2}$ \\ \hline
4 & 1 & 0.5 & 5 & 2.5 & 300 & 7.2 & 90 & 3000 & 8 \\ \hline
&  &  &  &  & 100 & 7.2 & 30 & 300 & 8 \\ \hline
&  &  &  &  & 30 & 7.2 & 7 & 20 & 7 \\ \hline
4 & 1 & 0.5 & 5 & 2.7 & 300 & 6.4 & 30 & 1000 & 0.8 \\ \hline
&  &  &  &  & 100 & 6.4 & 10 & 100 & 1.0 \\ \hline
&  &  &  &  & 30 & 6.4 & 3 & 9 & 0.9 \\ \hline
&  &  &  &  & 10 & 6.4 & 1 & 1 & 0.9 \\ \hline
4 & 1 & 0.2 & 7 & 3 & 300 & 6.8 & 400 & 10000 & 900 \\ \hline
5 & 1.5 & 0.2 & 7 & 3.5 & 300 & 7.4 & 20 & 900 & 10 \\ \hline
&  &  &  &  & 100 & 7.4 & 8 & 90 & 15
\end{tabular}
\end{table}

\end{document}